\documentclass[journal]{IEEEtran}
\ifCLASSINFOpdf

\else
 
\fi

\usepackage{cite}
\usepackage{amsmath}
\usepackage{amssymb}
\usepackage{amsfonts}
\usepackage{graphicx}
\usepackage{multirow}
\usepackage{balance}
\usepackage{subcaption}
\usepackage{booktabs}
\usepackage{colortbl}

\begin{document}

\title{Advanced Assessment of Stroke in Retinal Fundus Imaging with Deep Multi-view Learning}

\author{Aysen Degerli, 
        Mika Hilvo,
        Juha Pajula, 
        Petri Huhtinen,
        and Pekka Jäkälä
\thanks{A. Degerli, M. Hilvo, and J. Pajula are with VTT Technical Research Centre of Finland, Tampere and Espoo, Finland (e-mail: name.surname@vtt.fi). }
\thanks{Petri Huhtinen is with Optomed Oyj, Oulu, Finland (e-mail: petri.huhtinen@optomed.com).}
\thanks{Pekka Jäkälä is with Kuopio University Hospital, Kuopio, Finland (e-mail: pekka.jakala@pshyvinvointialue.fi).}

}

\markboth{Journal of \LaTeX}{}

\maketitle

\begin{abstract}
Stroke is globally a major cause of mortality and morbidity, and hence accurate and rapid diagnosis of stroke is valuable. Retinal fundus imaging reveals the known markers of elevated stroke risk in the eyes, which are retinal venular widening, arteriolar narrowing, and increased tortuosity. In contrast to other imaging techniques used for stroke diagnosis, the acquisition of fundus images is easy, non-invasive, fast, and inexpensive. Therefore, in this study, we propose a multi-view stroke network (MVS-Net) to detect stroke and transient ischemic attack (TIA) using retinal fundus images. Contrary to existing studies, our study proposes for the first time a solution to discriminate stroke and TIA with deep multi-view learning by proposing an end-to-end deep network, consisting of multi-view inputs of fundus images captured from both right and left eyes. Accordingly, the proposed MVS-Net defines representative features from fundus images of both eyes and determines the relation within their macula-centered and optic nerve head-centered views. Experiments performed on a dataset collected from stroke and TIA patients, in addition to healthy controls, show that the proposed framework achieves an AUC score of 0.84 for stroke and TIA detection.
\end{abstract}


\IEEEpeerreviewmaketitle

\section{Introduction}
\IEEEPARstart{S}{troke} is a severe and prevalent manifestation of cerebrovascular disease that occurs due to an occluded or ruptured artery supplying blood to the brain \cite{PORTEGIES2016239}. World Health Organization (WHO) recognizes stroke as the second leading cause of death and disability \cite{campbell2019ischaemic}. According to the statistics reported by the World Stroke Organization (WSO), annually $12.2$ million new strokes occur, and $6.5$ million deaths are caused by stroke \cite{feigin2022world}. Although age is the most significant demographic risk factor, $16\%$ of strokes are encountered at the age of $15-49$, resulting in mortality and morbidity due to stroke \cite{feigin2022world, doi:10.1161/01.STR.28.7.1507}. Henceforth, the global cost for the diagnosis and treatment of stroke sharply increases as it is no longer confined to be primarily observed in the elderly population. Therefore, accurate and immediate diagnosis of stroke plays a significant role in the acceleration of its treatment and prevention of mortality and disability.

Cerebral vasculature is formed by the vessels maintaining the blood supply and flow to the brain \cite{ABSHER2002733}. Cerebrovascular disease (CVD) encompasses disorders or diseases of the cerebral blood vessels, blood flow, and oxygenation \cite{ABSHER2002733, andrade2012systematic}. Stroke is the most severe clinical manifestation among CVDs \cite{SHARMA2017455}. The main types of stroke are categorized as ischemic stroke, hemorrhagic stroke, and transient ischemic attack (TIA). $71\%$ of strokes are ischemic strokes that cause infarction in the brain, spinal cord, or retina \cite{campbell2019ischaemic}. Contrary to ischemic stroke, hemorrhagic stroke is a rare but more vital case that occurs when the artery supplying blood to the brain ruptures \cite{unnithan2022hemorrhagic}. According to the stroke definition of WHO, ischemic stroke symptoms continue over $\geq24$ hours or even lead to death \cite{hatano1976experience}. In comparison to ischemic stroke, TIAs last less than $24$ hours and resolve before leaving any permanent damage \cite{amarenco2020transient}. Hence, TIA is called a mini-stroke caused by the temporary interruption of blood flow to the brain \cite{johnston2002transient}. Of note, TIA results in a $20-25\%$ risk of future ischemic stroke as the pathogenesis of TIA and ischemic stroke is identical \cite{amarenco2020transient}. However, discriminating TIA from non-cerebrovascular diseases is a challenging task, and $30\%-50\%$ of patients are misdiagnosed \cite{vuong2015ocular}. Therefore, the diagnosis of TIA is vital to prevent a possible future ischemic stroke.

Pre-hospital stroke scale tests, electroencephalography (EEG), and imaging techniques are used in the initial diagnosis of a stroke. Accordingly, pre-hospital stroke scale tests such as the National Institutes of Health Stroke Scale (NIHSS), Face Arm Speech Time (F.A.S.T.), and Cincinnati Pre-hospital Stroke Scale (CPSS) are primary tests for detecting the presence of a stroke, which are used to assess the facial expressions, motor skills, and any speech disorder \cite{alijanpour2021different}. However, these tests face challenges due to the heterogeneous nature of stroke, which may result in unreliability, bias, and inability to manage time in diagnosing a stroke \cite{doi:10.1161/01.STR.30.8.1534,doi:10.1111/j.1747-4949.2009.00294.x}. On the other hand, EEG enables monitoring of the electrical activity of the brain, which provides valuable information for the diagnosis of various neurological disorders, including stroke. However, EEG signals are interpreted more efficiently in stroke diagnosis when utilized together with other imaging modalities \cite{chaddad2023electroencephalography}. Hence, imaging techniques are crucial for the initial stroke identification, since brain infarction is typically monitored using Computed Tomography (CT) or Magnetic Resonance Imaging (MRI) \cite{birenbaum2011imaging}. Considering the time, limited sensitivity, and cost of acquiring conventional CT and MRI images and radiation exposure in CT imaging, alternative imaging modalities are of great interest in the biomedical field \cite{10.1159/000362719, muir2006imaging}. 

Retina can reflect the elevated risk of stroke as it shares anatomical, embryological, and physiological similarities with the brain \cite{CHEUNG201789}. Hence, retinal fundus imaging is considered a potential biomarker of CVDs as a complementary imaging technique for stroke diagnosis. Fundus images are acquired by a fundus camera, which is a portable and non-invasive device generally used by ophthalmologists \cite{mishra2022fundus}. Accordingly, retinal arterial abnormalities associated with cerebral infarction and hemorrhage are revealed with retinal fundus imaging \cite{jeena2019retina, henderson2011hypertension}. These abnormalities include arteriolar fibrous thickening near the optic disc, retinal venular widening, increased tortuosity, arteriolar narrowing, and cotton wool spots \cite{CHEUNG201789, henderson2011hypertension}. Moreover, transient or permanent vision loss is a major outcome of stroke that can be caused by retinal ischemia \cite{pula2017eyes}. Therefore, retinal imaging is strongly suggested as a diagnostic tool in the early prediction of stroke \cite{jeena2019retina}.

In healthcare, deep learning is extensively used over medical imaging data, where computer-aided diagnosis of various diseases and medical emergency conditions is performed achieving state-of-the-art results. Deep learning-based approaches with Convolutional Neural Networks (CNNs) are widely used in stroke imaging including applications such as prediction, treatment, decision making, and segmentation of lesion regions in the brain \cite{ZHU2022147}. In the literature, many studies\cite{pachade2022detection, jeena2021comparative, jeena2019retina, jeena2019stroke, Lim_Lim_Xu_Ting_Wong_Lee_Hsu_2019, coronado2021towards, diagnostics12071714} have performed ischemic stroke detection using retinal fundus images with deep learning. Pachade et al. \cite{pachade2022detection} compared the performance of hand-crafted and self-supervised deep learning features extracted from fundus images in stroke detection, and achieved an Area Under Curve (AUC) of 0.57 using the k-Nearest Neighbors classifier. Jeena et al. proposed a stroke detection scheme by extracting various hand-crafted features in their several studies \cite{jeena2021comparative, jeena2019stroke, jeena2019retina}. In another study \cite{Lim_Lim_Xu_Ting_Wong_Lee_Hsu_2019}, authors evaluated a state-of-the-art CNN model to predict ischemic stroke using fundus images. Moreover, transfer learning has been performed in studies \cite{coronado2021towards, diagnostics12071714}, where pre-trained deep networks have been trained further with retinal fundus images, yielding AUC scores of 0.67 and 0.78 by Coronado et al. \cite{coronado2021towards} and Khan et al. \cite{diagnostics12071714}, respectively. Despite the various solutions proposed in the literature for stroke detection, the aforementioned studies use retinal fundus images collected from one eye only, which may not be clinically feasible. Moreover, there is still a gap for performance improvement with deep learning since studies have reported AUC values below $0.8$. In addition, previous studies have used retinal fundus images collected only from ischemic stroke, whereas other types of strokes are also significant to be included in the analysis for a reliable prediction.

In order to address the aforementioned issues, in this study, we propose a novel network, multi-view stroke network (MVS-Net) for assessing dual views of retinal fundus images captured from both eyes. Accordingly, we formulate a multi-class classification task to discriminate TIA, stroke, and healthy controls. MVS-Net consists of four input layers, where each layer takes separately dual views (macula-centered and optic nerve head-centered) of left and right eyes to assess a broader retinal region for stroke diagnosis as depicted in Fig. \ref{fig:framework}. The proposed MVS-Net is evaluated over the collected Stroke-Data dataset, which includes retinal fundus images from 73 stroke and 26 TIA patients, and  121 healthy controls. The collected Stroke-Data is a pioneering dataset, which includes for the first time retinal fundus images of transient ischemic attack (TIA) patients in addition to stroke and healthy controls. To the best of our knowledge, our study utilizes for the first time macula-centered and optic nerve head-centered views of retinal fundus images from both eyes of patients in stroke identification. Our solution supports the standardization of stroke diagnosis, where the proposed algorithm can easily be adapted to portable fundus cameras to be utilized in medical emergency conditions.

\section{Materials and Methods}

\subsection{Dataset}
The dataset used in this study was collected at Oulu University Hospital (OUH) and Kuopio University Hospital (KUH) in Finland between the years $2021$ and $2022$. The study was conducted under the Stroke-Data project that was approved by the ethics committee of North Ostrobothnia Hospital District ($22/2021$), which is the regional medical research ethics committee of the Wellbeing Services County of North Ostrobothnia. Moreover, the Finnish Medicines Agency Fimea approved the clinical trial for medical devices and performance evaluations. The ground truth of the Stroke-Data dataset was provided by neurologists who assessed the status of the patients during their stay at the hospital and performed the diagnosis of stroke and TIA with clinical evidence and other sensing and imaging modalities, i.e. CT and MRI. The types of stroke included in the dataset are ischemic stroke, hemorrhagic stroke, and TIA. Only one hemorrhagic stroke case was collected during data recruitment due to its rare condition compared to ischemic stroke. Hence, we refer to patients in stroke group having ischemic or hemorrhagic stroke and the symptoms of these types of strokes are permanent compared to TIA.

Table \ref{tab:characteristics} gives the characteristics of the study cohort, which consists of $220$ participants including $26$ TIA and $73$ stroke patients accompanied by $121$ healthy controls. The table reveals that the Stroke-Data dataset consists of $119$ female and $101$ male participants. In the data, the mean age varies from $57$ to $70$ on average in the groups. The age range in controls is the largest ($21-86$), while in the TIA and stroke groups, it is more limited ($31-85$ and $38-79$, respectively). During dataset collection, demographic characteristics were utilized when collecting controls to form a balanced distribution across groups to avoid any bias in these respects in deep learning model training.

\begin{table*}[t!]
\centering
\caption{The characteristics of the study cohort, where $n$ indicates the number of subjects.}
\resizebox{\linewidth}{!}{
\begin{tabular}{llcccccc}
& & \multicolumn{2}{c}{Control} & \multicolumn{2}{c}{Stroke} & \multicolumn{2}{c}{TIA}
\\ \cmidrule(lr){3-4}\cmidrule(lr){5-6}\cmidrule(lr){7-8}
             &          & \begin{tabular}[c]{@{}c@{}}Female \\ $(n=82)$ \end{tabular}  & \begin{tabular}[c]{@{}c@{}}Male \\ $(n=39)$ \end{tabular}  & \begin{tabular}[c]{@{}c@{}}Female \\ $(n=31)$ \end{tabular}  & \begin{tabular}[c]{@{}c@{}}Male \\ $(n=42)$ \end{tabular}  &  \begin{tabular}[c]{@{}c@{}}Female \\ $(n=6)$ \end{tabular} &  \begin{tabular}[c]{@{}c@{}}Male \\ $(n=20)$ \end{tabular} \\ \toprule
Age \\(years)  
             & Mean     & $57$    & $59$      & $66$     & $65$     & $63$    & $70$ \\
             & Min$-$Max  & $23-86$    & $21-83$      & $31-85$     & $44-82$     & $46-79$    & $38-86$ \\
             &          &         &           &          &          &         &      \\
Weight \\(kg)  & Mean   & $73.5$  & $81.9$    & $76.4$   & $84$     & $71.3$  & $90.3$ \\
             & Min$-$Max  & $41-138$    & $65-104$      & $42-148$     & $56-118$     & $64-75$    & $75-105$    \\
             &          &         &           &          &          &         &         \\
Height \\(cm) & Mean    & $163.9$  & $177.1$  & $162.6$  & $177.5$  & $161.3$ & $176.9$ \\
            & Min$-$Max   & $150-178$    & $171-188$    & $150-176$    & $168-190$    & $153-169$   & $166-190$    \\
            &           &          &          &          &          &          &          \\
Education level \\(number) & Primary education              & $14$ & $13$ & $9$  & $11$ & $2$ & $7$  \\
                           & Upper secondary education      & $25$ & $9$  & $15$ & $20$ & $2$ & $9$ \\ 
                           & Bachelors or equivalent        & $26$ & $11$ & $6$  & $4$  & $1$ & $2$  \\
                           & Masters/Doctorate or equivalent & $17$ & $6$  & $1$  & $7$  & $1$ & $2$  \\
                           &   &     &     &          &          &     &    \\
Work Status \\(number)   & Employee                & $47$ & $16$ & $9$  & $12$ & $2$ & $5$  \\
                         & Entrepreneur            & $1$  & $2$  & $2$  & $6$  & $-$ & $2$  \\
                         & Retired                 & $32$ & $20$ & $19$ & $24$ & $4$ & $13$ \\
                         & Student                 & $1$  & $-$  & $-$  & $-$  & $-$ & $-$  \\
                         & Unable to work         & $-$  & $-$  & $1$  & $-$  & $-$ & $-$  \\ 
                         & On parental leave       & $1$  & $-$  & $-$  & $-$  & $-$ & $-$  \\ 
                         & Information unavailable & $-$  & $1$  & $-$  & $-$  & $-$ & $-$  \\ 
                         
                         \bottomrule
\end{tabular}}
\label{tab:characteristics}
\end{table*}

Research study nurses acquired retinal fundus images using the Optomed Aurora IQ fundus camera at the Stroke Units of hospitals. The most important camera settings were auto-focus and auto-exposure. The study-specific protocol was provided by the manufacturer to capture the images. Accordingly, retinal fundus images from both eyes of the participants were captured. From each eye, dual views were obtained, which refer to macula-centric and optic nerve head-centric images that visualize the central visual field and papilla in the middle, respectively. In the Stroke-Data dataset, a total of $802$ retinal fundus images were collected from the participants. Table \ref{tab:dataset} shows the details of the dataset, where the number of images from each view of the eye is not equal. The reason is missing data, due to low quality images or difficulties in imaging during the data collection phase, such as poor health conditions of the participants. For the cases in which data quality is unacceptable to analyze, research study nurses collected multiple images (at most three times) before moving to the next eye or view. Accordingly, the data collection phases were as follows:
\begin{enumerate}
    \item The test protocol was introduced to the participant by the research study nurse.
    \item The research study nurse described the basic characteristics of the participant's fundus by visual inspection.
    \item The research study nurse captured two images from both eyes that corresponded to macula-centric and optic nerve head-centric retinal fundus images.
    \item After the recording session was finished, the research study nurse transferred images from the camera's memory card to the research computer for storage.
    \item The research study nurse cleaned the camera device for the next use.
\end{enumerate}
The measurement protocol took approximately $5$ minutes, which is similar to shooting with a regular digital camera. In the data collection phase, the user interface of the fundus camera guided the research study nurses, where every captured image was displayed before the nurse decided either to store the image or capture a new one depending on the image quality.

For the analysis, from each subject, a set of four images are gathered, which are resized to $224\times224$ to align the input dimensions of deep network topologies. The dataset is formed in a way that in case of a missing image from one view, it is duplicated by flipping the image of that specific view from the contrary eye. However, in case of missing images of the same view from both eyes, the subject is excluded from the dataset due to having no corresponding view from the other eye.

\begin{table}[t!]
\centering
\caption{Details of the dataset; number of images with respect to class, eye-side, and view.}
\renewcommand{\arraystretch}{1.1}
\resizebox{\linewidth}{!}{
\begin{tabular}{cccccc}
\toprule
\multirow{2}{*}{\textbf{Class}} & \multirow{2}{*}{\textbf{\#participants}} & \multicolumn{2}{c}{\textbf{Right eye}} & \multicolumn{2}{c}{\textbf{Left eye}} \\
        &     & \textit{View 1} & \textit{View 2} & \textit{View 1} & \textit{View 2} \\ \midrule
Control & $121$ & $109$ & $120$ & $110$ & $117$ \\ 
Stroke  & $73$  & $56$  & $71$  & $59$  & $69$  \\
TIA     & $26$  & $18$  & $26$  & $22$  & $25$  \\ \midrule
\textbf{Total}  & $220$ & $183$ & $217$ & $191$  & $211$ \\
\end{tabular}}
\label{tab:dataset}
\end{table}

\begin{figure*}[t!]
    \centering
    \includegraphics[width=.95\linewidth]{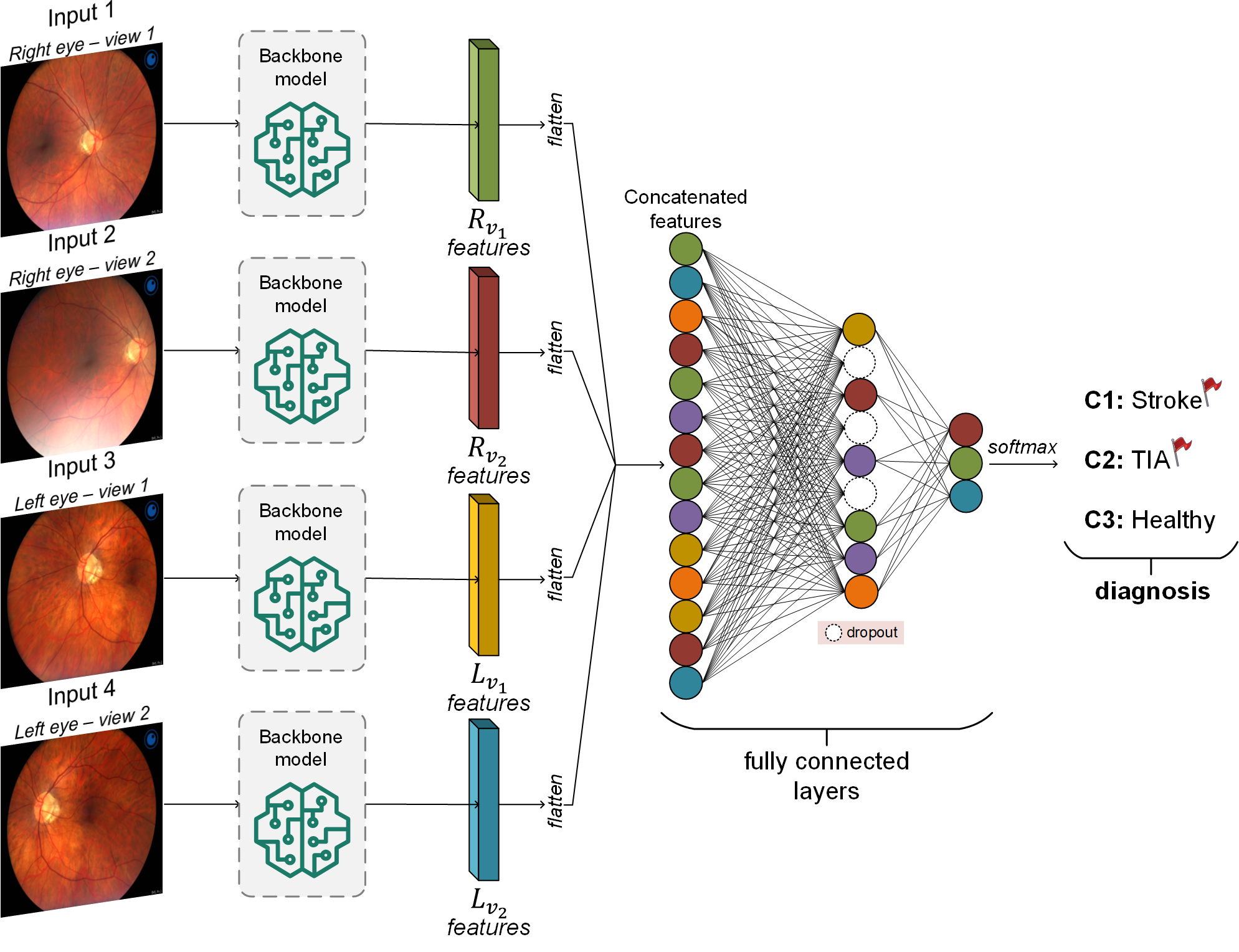}
    \caption{The proposed multi-view stroke network (MVS-Net) is illustrated, where dual-view retinal fundus images are given to MSV-Net as inputs to predict stroke, TIA, and healthy controls.}
    \label{fig:framework}
\end{figure*}

\subsection{Multi-view Stroke Network}
Multi-view learning (MVL) is an emerging technique that is used to exploit complementary information of multiple modalities by projecting data from different views into a common feature space \cite{YAN2021106}. In the literature, deep MVL is generally deployed with CNNs, where high-level feature representations are extracted from the multiple views of data and integrated in order to map features into a target. This study proposes a novel deep multi-view learning network for the identification of stroke using retinal fundus images. Initially, we formulate the problem as a binary classification task to detect any presence of stroke, where the positive class includes stroke and TIA patients. In addition, we set an additional formulation, where we perform a multi-class classification task to discriminate stroke, TIA, and healthy controls. Accordingly, we focus on the initial setup, which is essential in medical emergency cases to detect any stroke presence directly. However, the latter problem formulation is used to investigate the discrimination of TIA patients from ischemic stroke.

\subsubsection{Model Structure}
Let us denote the set of images as $\mathbf{X} \in \mathbb{R}^{N\times w\times h \times3}$, where $N$ is the number of images in the set, $w$ and $h$ are the width and height of images, respectively. A set of four retinal fundus images $\mathbf{X}=\{\mathbf{I}_{R_{v_1}}, \mathbf{I}_{R_{v_2}},\mathbf{I}_{L_{v_1}},\mathbf{I}_{L_{v_2}}\}$ is gathered from each subject for the analysis, which includes optic nerve head-centric ($v_1$) and macula-centric ($v_2$) views of both the left, $L$ and right, $R$ eye of the participants and their corresponding ground truth labels $\mathbf{y} \in \{0, 1, 2\}$. The proposed MVS-Net maps the set of input images $\mathbf{X}$ to the predicted class, $\hat{\mathbf{y}}:\hat{\mathbf{y}} \leftarrow \Upsilon
_{\beta, \iota}(\mathbf{X}, \mathbf{y})$, where $\beta$ represents the selected backbone model and $\iota$ is the fully connected layers as depicted in Fig. \ref{fig:framework}. Finally, the proposed network is trained over the dataset $\mathbf{D}=\{\mathbf{X}_s, \mathbf{y}_s\}^S_{s=1}$, where $S$ is the number of image sets.

\begin{table*}[t!]
\centering
\caption{Average stroke and TIA detection performance results (\%) of MVS-Net model computed from 5-folds, where the highest scores are highlighted in \textbf{bold}.}
\resizebox{.77\linewidth}{!}{
\begin{tabular}{lcccccc}
\toprule
\multicolumn{1}{c}{Backbone Model} & \multicolumn{1}{c}{Sensitivity} & \multicolumn{1}{c}{Specificity} & \multicolumn{1}{c}{Precision} & \multicolumn{1}{c}{F1-Score} & \multicolumn{1}{c}{AUC} & \multicolumn{1}{c}{Accuracy} \\ 
\midrule
 BiT & $0.606$ & $0.744$ & $0.659$ & $0.632$ & $0.786$ & $0.682$ \\

 DenseNet-121 & $0.596$ & $0.785$ & $0.694$ & $0.641$ & $0.778$ & $0.700$ \\
 
 Inception-v3 & $0.626$ & $0.810$ & $0.729$ & $0.674$ & $0.800$ & $0.727$ \\

 ResNet50 & $\textbf{0.687}$ & $0.785$ & $0.723$ & $0.705$ & $\textbf{0.837}$ & $0.741$ \\
 
 VGG19 & $0.515$ & $0.736$ & $0.615$ & $0.560$ & $0.711$ & $0.636$ \\

 Xception & $0.636$ & $\textbf{0.868}$ & $\textbf{0.798}$ & $\textbf{0.708}$ & $0.803$ & $\textbf{0.764}$ \\

\bottomrule 
\end{tabular}}
\label{tab:experiments_1}
\end{table*}

The proposed MVS-Net consists of multi-input channels each corresponding to a retinal fundus image, $\mathbf{I}$ in the set of images, $\mathbf{X}$. Several state-of-the-art CNNs are used as backbone models, $\beta$ to extract high-level features from each view to map those into the predicted class label. Each input channel is connected to a backbone model for high-level feature extraction, where Densenet-121 \cite{huang2017densely}, Inception-v3 \cite{szegedy2016rethinking}, ResNet50 \cite{he2016deep}, VGG19 \cite{simonyan2015a}, Xception \cite{8099678}, and BigTransfer \cite{10.1007/978-3-030-58558-7_29} (BiT) are used due to their state-of-the-art performance in many computer vision tasks. The selected models were pre-trained over the ImageNet \cite{5206848} dataset, which is advantageous for initializing the weights for faster convergence during training. Features are extracted with backbone models $\mathbf{f} \leftarrow \beta(.)$ from each input retinal fundus image, and concatenated after $\mathbf{f}$ is vectorized and downsampled by global average pooling layer as $\mathbf{F} = [\mathbf{f}_{R_{v_1}},\mathbf{f}_{R_{v_2}},\mathbf{f}_{L_{v_1}},\mathbf{f}_{L_{v_2}}], \mathbf{f} \in \mathbb{R}^{d}$, where $d$ is the feature dimension after flattening. Then, a fully connected layer with \textit{rectified linear unit} (ReLU) activation function, $\delta$ is attached as $\delta(\iota(\mathbf{F}))$ for classification. The neuron size of the fully connected layer is dependent on the feature dimension, which is determined as $\frac{d}{6}$. After the fully connected layer, a dropout layer with a rate of $0.4$ is attached, which randomly drops units from MVS-Net to prevent overfitting. Lastly, a fully connected layer with $3$-neurons using \textit{softmax} activation function is attached as the output of the model. In the case of the binary classification setup, where the aim is to detect stroke, the output size is set to $2$-neurons.

\subsubsection{Model Training}
The model training is performed by adapting a stratified $5$-fold cross-validation scheme, where training and test sets are separated with a ratio of 80\% and 20\%, respectively. The formation of training and test sets are performed patient-wise. During the training, we perform data augmentation in the training set by augmenting images in dual view of each eye up to $1000$ samples for each class, which corresponds to $12$K images in the training set in total. Accordingly, for the data augmentation, we shift images horizontally and vertically, flip, apply random shear, rotate images randomly between $0-180$ degrees, and zoom with a range of $10$. The model is trained with a batch size of $32$ using Adam optimizer with the learning rate of $10^{-5}$ over $300$ epochs. Early stopping with $5$ epochs of patience is applied to prevent overfitting. In addition, the weights of the model are saved at the epoch where the categorical accuracy hits the highest. The loss function is determined as categorical cross-entropy for stroke identification. 

\begin{figure}[b!]
         \centering
         \begin{subfigure}[t]{.48\linewidth}
         {\includegraphics[width=\linewidth]{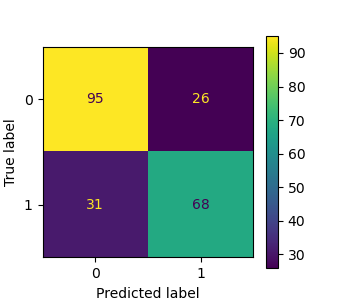}}
         \caption{}
         \end{subfigure}
         \centering
         \begin{subfigure}[t]{.48\linewidth}
         {\includegraphics[width=\linewidth]{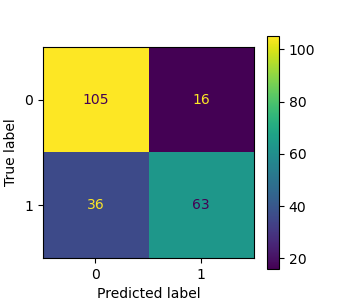}}
         \caption{}
         \end{subfigure}      
         \centering
         \begin{subfigure}[t]{.48\linewidth}
         {\includegraphics[width=\linewidth]{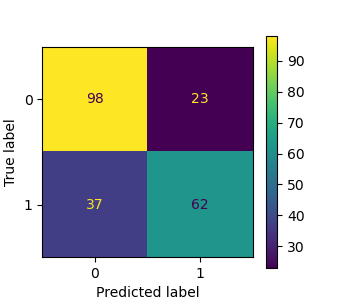}}
         \caption{}
         \end{subfigure} 
     \centering
     \caption{Confusion matrices of the MVS-Net, where the backbone is (a) ResNet50, (b) Xception, and (c) Inception-v3 models.}
     \label{fig:confusion_matrices}
\end{figure}

\subsubsection{Evaluation Metrics}
The performance is evaluated using metrics calculated from the accumulated confusion matrix of folds, which consists of true positive (TN), true negative (TN), false positive (FP), and false negative (FN) elements. In experimental results, we report the following metrics with a cut-off value of $0.5$: \textit{sensitivity} measured by $\frac{\text{TP}}{\text{TP}+\text{FN}}$, \textit{specificity} defined as $\frac{\text{TN}}{\text{TN}+\text{FP}}$, \textit{precision} calculated by $\frac{\text{TP}}{\text{TP}+\text{FP}}$, \textit{F1-Score} is the harmonic average between sensitivity and precision, \textit{accuracy} measured by $\frac{\text{TP}+\text{TN}}{\text{TP}+\text{TN}+\text{FP}+\text{FN}}$. Lastly, the area under curve (AUC) score is measured from the Receiver Operating Characteristics (ROC), which is the plot of true positive rate (sensitivity) and false positive rate ($1-\text{specificity}$) values as changing the threshold of the predictions.

\begin{table}[b!]
\centering
\caption{Number of parameters in MVS-Net as the backbone changes and trained for stroke detection (binary classification), where the highest number of parameters are highlighted in \textbf{bold}.}
\renewcommand{\arraystretch}{1.1}
\resizebox{\linewidth}{!}{
\begin{tabular}{lccc}
\toprule
\multicolumn{1}{c}{Backbone Model} & \multicolumn{1}{c}{Trainable} & \multicolumn{1}{c}{Non-trainable} & \multicolumn{1}{c}{Total}  \\ 
\midrule
BiT & $11,186,177$ & $23,500,352$ & $34,686,529$ \\
DenseNet-121 & $9,749,376$ & $83,648$ & $9,833,024$ \\
Inception-v3 & $32,954,529$ & $34,432$ & $32,988,961$ \\
ResNet50 & $34,720,769$ & $53,120$ & $\textbf{34,773,889}$ \\
VGG19 & $20,723,777$ & $0$ &  $20,723,777$ \\
Xception & $31,993,129$ & $54,528$ & $32,047,657$ \\
\bottomrule 
\end{tabular}}
\label{tab:model_parameters}
\end{table}

\begin{table*}[t!]
\centering
\caption{Average performance (\%) of MVS-Net model reported for each class computed from 5-folds, where the highest scores are highlighted in \textbf{bold}.}
\resizebox{.8\linewidth}{!}{
\begin{tabular}{clcccccc}
\toprule
Class & \multicolumn{1}{c}{Backbone Model} & \multicolumn{1}{c}{Sensitivity} & \multicolumn{1}{c}{Specificity} & \multicolumn{1}{c}{Precision} & \multicolumn{1}{c}{F1-Score} & \multicolumn{1}{c}{AUC} & \multicolumn{1}{c}{Accuracy} \\ 
\midrule

 \multirow{6}{*}{\textit{Control}} & BiT	& $0.802$ & $0.546$ & $0.683$ & $0.738$ & $0.763$ & $0.627$ \\

 & DenseNet-121 & $0.777$ & $0.535$ & $0.671$	& $0.720$ & $0.761$ & $0.609$ \\

 & Inception-v3	& $0.777$ & $0.586$ & $0.696$ & $0.734$ & $0.786$ & $0.614$ \\
 
 & ResNet50 & $0.793$ & $\textbf{0.596}$ & $\textbf{0.706}$ & $\textbf{0.747}$ & $\textbf{0.803}$ & $\textbf{0.646}$ \\

 & VGG19 & $0.711$ & $0.475$ & $0.623$ & $0.664$ & $0.691$ & $0.536$ \\

 & Xception & $\textbf{0.818}$ & $0.535$ & $0.683$ & $0.744$ & $0.771$ & $0.591$ \\ \hline

 \multirow{6}{*}{\textit{Stroke}} &  BiT & $0.507$ & $0.796$ & $0.552$ & $0.529$ & $0.738$ & $0.627$ \\

 & DenseNet-121 & $0.521$ & $0.810$ & $0.576$ & $0.547$ & $\textbf{0.769}$ & $0.609$ \\
 
 & Inception-v3 & $0.493$ & $0.769$ & $0.514$ & $0.504$ & $0.739$ & $0.614$ \\

 & ResNet50 & $\textbf{0.575}$ & $0.803$ & $\textbf{0.592}$ & $\textbf{0.583}$ & $0.766$ & $\textbf{0.646}$ \\

 & VGG19 & $0.397$ & $0.742$ & $0.433$ & $0.414$ & $0.636$ & $0.536$ \\

 & Xception & $0.397$ & $\textbf{0.830}$ & $0.537$ & $0.457$ & $0.737$ & $0.591$ \\ \hline

 \multirow{6}{*}{\textit{TIA}} &  BiT & $0.154$ & $\textbf{0.964}$ & $\textbf{0.364}$ & $0.216$ & $0.699$ & $0.627$ \\

 & DenseNet-121 & $0.077$ & $0.938$ & $0.143$ & $0.100$ & $0.651$ & $0.609$ \\

 & Inception-v3 & $\textbf{0.192}$ & $0.949$ & $0.333$ & $\textbf{0.244}$ & $0.672$ & $0.614$ \\

 & ResNet50 & $0.154$ & $0.954$ & $0.308$ & $0.205$ & $\textbf{0.714}$ & $\textbf{0.646}$ \\

 & VGG19 & $0.115$ & $0.938$ & $0.200$ & $0.146$ & $0.629$ & $0.536$ \\

 & Xception & $0.077$ & $0.902$ & $0.095$ & $0.085$ & $0.630$ & $0.591$ \\

\bottomrule
\end{tabular}}
\label{tab:experiments_2}
\end{table*}

\begin{figure*}[t!]
         \centering
         \begin{subfigure}[t]{0.32\textwidth}        
         {\includegraphics[width=\textwidth]{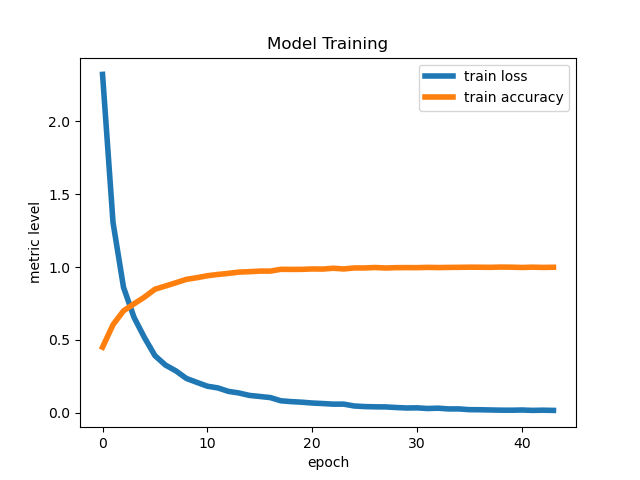}}
         \caption{BiT}
         \end{subfigure}                
         \hfill        
        \centering
         \begin{subfigure}[t]{0.32\textwidth}
         {\includegraphics[width=\textwidth]{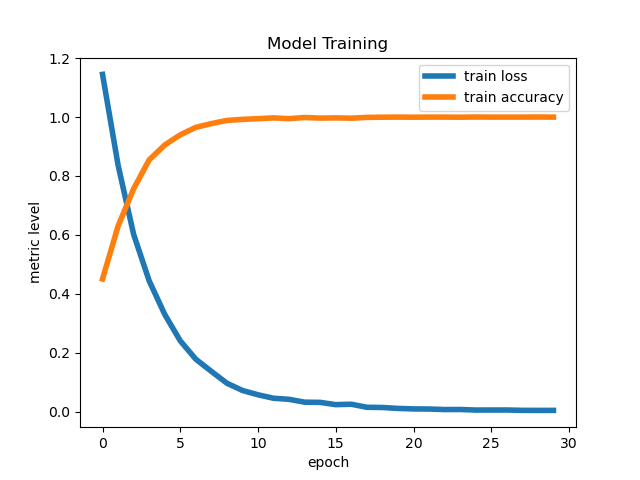}}
         \caption{DenseNet-121}
         \end{subfigure}                 
         \hfill        
        \centering
         \begin{subfigure}[t]{0.32\textwidth}
         {\includegraphics[width=\textwidth]{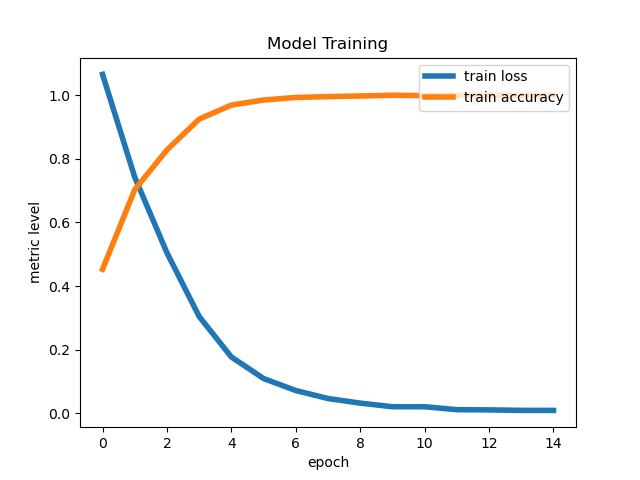}}
         \caption{Inception-v3}
         \end{subfigure}             
         \hfill
         \centering
         \begin{subfigure}[t]{0.32\textwidth}
         {\includegraphics[width=\textwidth]{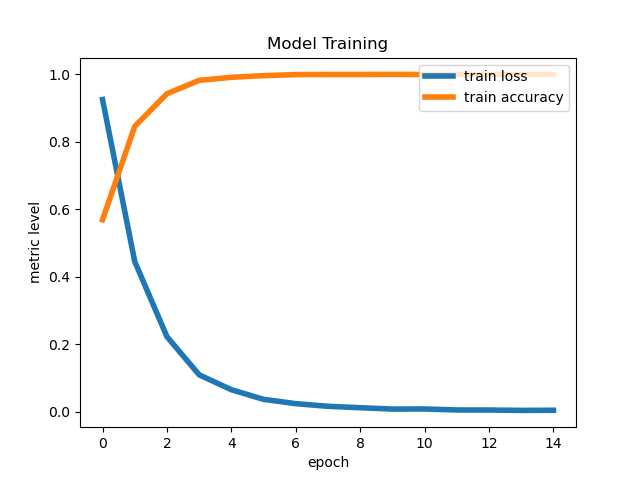}}
         \caption{ResNet50}
         \end{subfigure}          
         \hfill         
        \centering
         \begin{subfigure}[t]{0.32\textwidth}
         {\includegraphics[width=\textwidth]{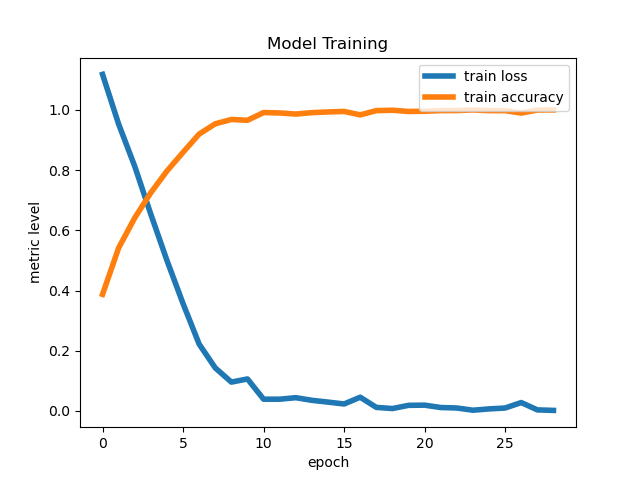}}
         \caption{VGG19}
         \end{subfigure}          
         \hfill        
        \centering
         \begin{subfigure}[t]{0.32\textwidth}
         {\includegraphics[width=\textwidth]{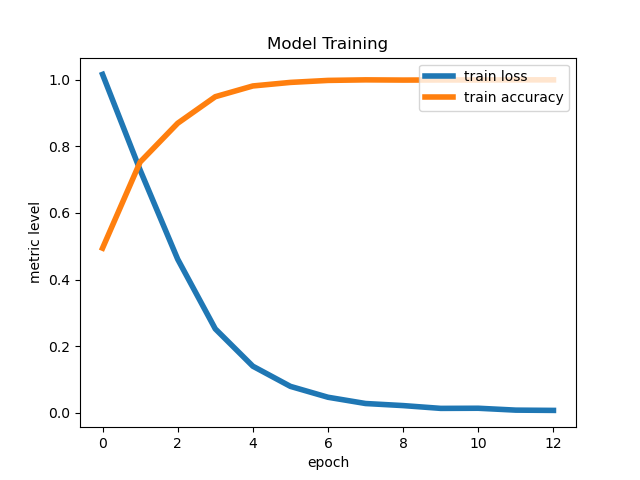}}
         \caption{Xception}
         \label{fig:training_b}
         \end{subfigure} 
     \centering
     \caption{The model training curves of the proposed MVS-Net are plotted with respect to train accuracy and loss, when the network is trained for discriminating stroke, TIA, and healthy controls. The backbone of the proposed model is indicated in the subfigure titles.}
     \label{fig:training}
\end{figure*}

\section{Results}
In this section, we report the performance of the proposed MVS-Net model for both binary and multi-class classification, where the latter case focuses on discriminating stroke, TIA, and healthy controls, whereas in binary classification the aim is to detect any stroke presence. Table \ref{tab:experiments_1} represents the results for stroke detection -  the binary classification case, where the performance of MVS-Net using different backbone models is reported. The proposed MVS-Net achieves the highest AUC score of $0.84$ and sensitivity level of $0.69$ with ResNet50 as the backbone model. On the other hand, the highest F1-Score of $0.71$ and accuracy of $0.76$ are achieved by the Xception backbone model. In Fig. \ref{fig:confusion_matrices}, confusion matrices of the backbone models with the top three AUC scores are provided, where the largest number of TNs are obtained with the Xception backbone model. Further investigation on model parameters reveals that the leading backbone model - ResNet50 has $34$M parameters considering both trainable and non-trainable parameters, which is the highest number given in Table \ref{tab:model_parameters}. On the contrary, the large number of network parameters obstruct its compatibility with small devices. 

\begin{table*}[t!]
\centering
\caption{Average performance (\%) of the \textit{single-view} approach computed from 5-folds.}
\resizebox{.9\linewidth}{!}{
\begin{tabular}{cclcccccc}
\toprule
Task & Class & \multicolumn{1}{c}{Backbone Model} & \multicolumn{1}{c}{Sensitivity} & \multicolumn{1}{c}{Specificity} & \multicolumn{1}{c}{Precision} & \multicolumn{1}{c}{F1-Score} & \multicolumn{1}{c}{AUC} & \multicolumn{1}{c}{Accuracy} \\ 
\hline

\multirow{6}{*}{\rotatebox[]{90}{{\textit{Binary}}}} &  & BiT & $0.528$ & $0.756$ & $0.629$ & $0.574$ & $0.699$ & $0.655$ \\

& & DenseNet-121 & $0.525$ & $0.766$ & $0.639$ & $0.576$ & $0.733$ & $0.660$ \\
 
& & Inception-v3 & $0.586$ & $0.714$ & $0.617$ & $0.601$ & $0.719$ & $0.658$ \\

& & ResNet50 & $0.392$ & $0.596$ & $0.433$ & $0.411$ & $0.513$ & $0.506$ \\
 
& & VGG19 & $0.603$ & $0.635$ & $0.565$ & $0.583$ & $0.664$ & $0.621$ \\

& & Xception & $0.581$ & $0.721$ & $0.620$ & $0.600$ & $0.695$ & $0.659$ \\ \hline

\multirow{18}{*}{\rotatebox[]{90}{{\textit{Multi-class}}}}& \multirow{6}{*}{\rotatebox[]{90}{Control}} & BiT	& $0.801$ & $0.308$ & $0.596$ & $0.683$ & $0.575$ & $0.553$ \\

& & DenseNet-121 & $0.786$ & $0.486$ & $0.661$ & $0.718$ & $0.728$ & $0.604$ \\

& & Inception-v3 & $0.795$ & $0.439$ & $0.643$ & $0.711$ & $0.691$ & $0.599$ \\
 
& & ResNet50 & $0.328$ & $0.694$ & $0.577$ & $0.418$ & $0.491$ & $0.406$ \\

& & VGG19 & $0.692$ & $0.544$ & $0.659$ & $0.675$ & $0.688$ & $0.556$ \\

& & Xception & $0.740$ & $0.519$ & $0.662$ & $0.699$ & $0.685$ & $0.587$ \\ \cline{2-9}

 & \multirow{6}{*}{\rotatebox[]{90}{Stroke}} &  BiT & $0.313$ & $0.793$ & $0.424$ & $0.361$ & $0.569$ & $0.553$ \\

 & & DenseNet-121 & $0.493$ & $0.7582$ & $0.498$ & $0.495$ & $0.708$ & $0.604$ \\
 
 & & Inception-v3 & $0.463$ & $0.784$ & $0.510$ & $0.485$ & $0.693$ & $0.599$ \\

 & & ResNet50 & $0.679$ & $0.316$ & $0.326$ & $0.441$ & $0.481$ & $0.406$ \\

 & & VGG19 & $0.489$ & $0.715$ & $0.455$ & $0.471$ & $0.668$ & $0.556$ \\

 & & Xception & $0.519$ & $0.720$ & $0.474$ & $0.496$ & $0.668$ & $0.587$ \\ \cline{2-9}

 & \multirow{6}{*}{\rotatebox[]{90}{TIA}} &  BiT & $0.011$ & $0.996$ & $0.250$ & $0.021$ & $0.498$ & $0.553$ \\

 & & DenseNet-121 & $0.022$ & $0.992$ & $0.250$ & $0.040$ & $0.616$ & $0.604$ \\

 & & Inception-v3 & $0.022$ & $0.990$ & $0.222$ & $0.040$ & $0.569$ & $0.599$ \\

 & & ResNet50 & $0.0$ & $1.0$ & $0.0$ & $0.0 $ & $0.528$ & $0.406$ \\

 & & VGG19 & $0.076$ & $0.942$ & $0.143$ & $0.099$ & $0.604$ & $0.556$  \\

 & & Xception & $0.022$ & $0.985$ & $0.154$ & $0.038$ & $0.541$ & $0.587$ \\

\bottomrule
\end{tabular}}
\label{tab:experiments_comparison}
\end{table*}

In this study, we train MVS-Net to discriminate stroke, TIA, and healthy controls with the multi-class classification setup by adjusting the output neuron size. Table \ref{tab:experiments_2} shows the results of the detection performance of MVS-Net model reported separately for each class in the dataset. For each class, ResNet50 backbone model generally achieves the highest AUC with $0.81$ score for healthy controls, $0.77$ for stroke, and $0.71$ for TIA. In fact, it can be depicted from Fig. \ref{fig:training} that ResNet50 backbone model has the fastest convergence, whereas VGG19 training includes fluctuations with a later convergence that results in the lowest performance. Experimental results reveal that discriminating TIA from the rest of the dataset is the most challenging, which was expected considering the difficulties that medical experts face in diagnosing TIA patients.

\textbf{Comparison to Existing Methods.} We investigated the performance of MVS-Net model when trained using single-view retinal fundus images. Accordingly, we treat each image individually by assigning the ground truth based on the patient's, where the confusion matrix is formed image-wise with $802$ elements, which is equal to the number of images in the dataset. For a fair comparison, we use the same experimental setup as in the multi-view approach. We adjust the MVS-Net for single-view by using only one input channel, which is attached to the backbone model for high-level feature extraction. The extracted feature $\mathbf{F}$ is connected to a fully connected layer, a dropout layer, and the output layer with the same parameters as in the MVS-Net when trained for the multi-view. In addition, the same model training setup is followed in the single-view experiments.

\begin{figure*}[t!]
         \centering
         \begin{subfigure}[t]{.4\linewidth}
         {\includegraphics[width=\linewidth]{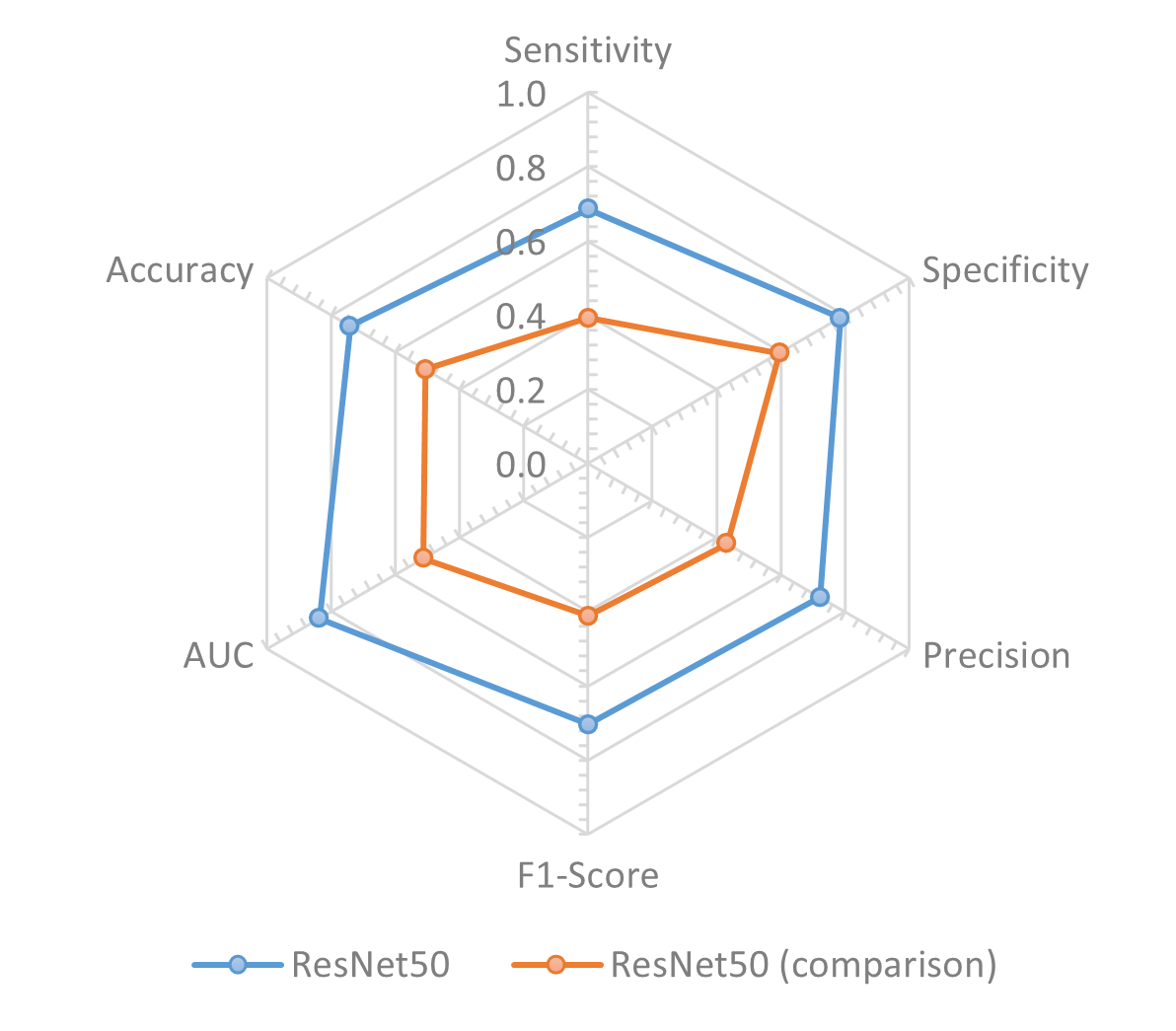}}
         \caption{}
         \end{subfigure}
         \centering
         \begin{subfigure}[t]{.4\linewidth}
         {\includegraphics[width=\linewidth]{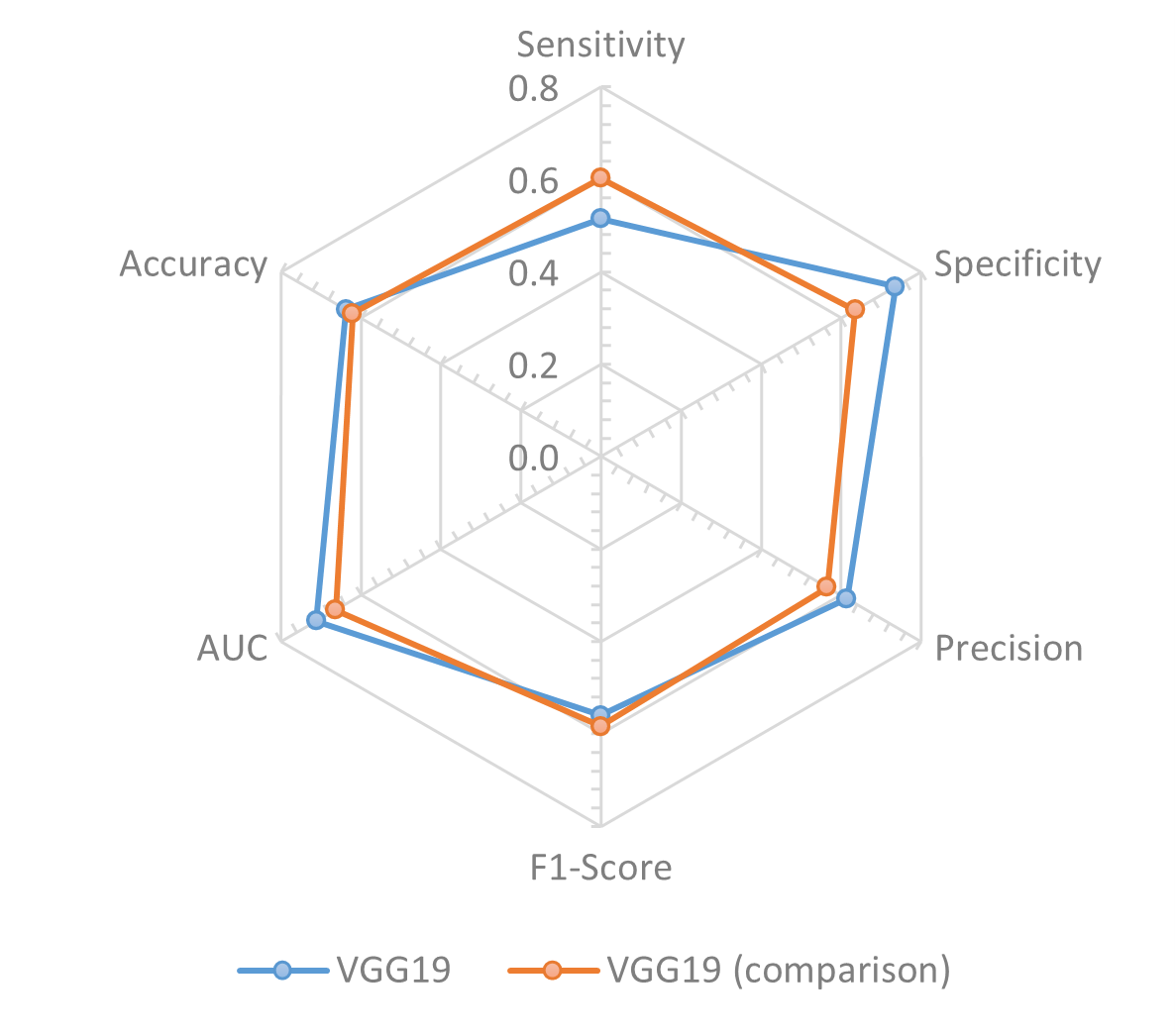}}
         \caption{}
         \end{subfigure}      
     \centering
     \caption{Radar charts of MVS-Net with ResNet50 (a) and VGG19 (b) backbone models for comparison to existing methods.}
     \label{fig:radar_charts}
\end{figure*}

Table \ref{tab:experiments_comparison} provides the performance of stroke detection, where AUC scores drop significantly when MVS-Net is trained using single-view retinal fundus images. The leading backbone model, ResNet50 degrades its sensitivity level by $29.52\%$ and the Xception backbone model has $0.18$ less score in the precision in single-view stroke detection for the binary classification setup. The provided radar charts in Fig. \ref{fig:radar_charts} illustrate the performance comparison between the proposed and existing approaches, where both the leading and lowest-performed backbone models, ResNet50 and VGG19 show a degraded performance level in the single-view approach. Moreover, the performance of the proposed MVS-Net decreases when it is trained to discriminate stroke, TIA, and healthy controls (multi-class) as reported in Table \ref{tab:experiments_comparison}. Hence, results reveal that detection from single-view retinal fundus images proposed by many studies in the literature is not a suitable approach for stroke detection. Therefore, this study for the first time shows that using multi-view retinal fundus images for stroke detection outperforms the single-view approach.

\section{Discussion}
Stroke diagnosis seeks new diagnostics that can be used in case of emergency to identify stroke symptoms and transfer patients to a hospital quickly. In this study, we propose a method to standardize stroke diagnosis with retinal fundus images that can be used even in remote areas, where urgent access to a hospital and neurologist is restricted. The proposed method can be used in a fundus camera, which is non-invasive and cheap, accelerating the first-aid detection of stroke with minimal subjectivity. 

The collected Stroke-Data dataset is unique in terms of having TIA patients contrary to studies in the literature performing deep learning-based stroke detection using retinal fundus images. In the literature, many studies have proposed approaches to the detection of stroke from single-view retinal fundus images, which may not be reliable in real-life case scenarios; hence, it may not be clinically feasible. The drawback of single-view stroke detection is that abnormalities caused by stroke may not be visible only in one eye. Moreover, even for the same eye, using dual views provides a broader region to be assessed for the diagnosis, which increases the reliability of the proposed methods. The multi-view collection of images from the same participants brings improvements to the methodology and detection. Therefore, retinal fundus imaging with deep learning is recommended as it has great low-cost potential for stroke treatment improvement. The proposed approach can also be used to compare other diseases that can be investigated with retinal fundus imaging and deep learning (like diabetes).

Our study has several limitations. The Stroke-Data dataset is quite small considering that it includes $802$ images from $220$ participants, which is not very suitable for training deep learning models. In order to overcome this issue, we applied data augmentation to increase the number of samples during training. We show that the proposed deep learning model can give promising results, even if the dataset is small for a proper solution. On the other hand, there are no available retinal fundus image datasets that have multi-views from each participant. Hence, it is not possible to test our model on another dataset. 

The definition of stroke includes several types and their subcategories \cite{sacco2013updated}. In this study, we have analyzed the main types of stroke, which are ischemic stroke, hemorrhagic stroke, and TIA. As hemorrhagic stroke is a rare condition compared to ischemic stroke, this study included only one hemorrhagic patient. In our analysis, we consider ischemic and hemorrhagic types in the stroke category as they present permanent symptoms compared to TIA patients. Including all types of strokes in the dataset is crucial for a reliable stroke detection approach; however, its feasibility in the data collection phase is challenging. 

The number of TIA patients in the dataset is much less compared to other classes which makes the model harder to discriminate TIA. Considering the fact that TIA detection is even more challenging for medical doctors, it is also reflected in the performance of our proposed model which achieves the lowest performance level. On the other hand, the proposed method achieves approximate levels of AUC for stroke and TIA. Hence, the default threshold value may not be optimal for the TIA cases and it can be changed to achieve more balanced sensitivity/specificity levels. In the end, TIA might require a dedicated approach as it might not be detected in the same way as stroke.

\section{Conclusion}
This study proposes for the first time a multi-view deep learning network for stroke detection using dual views (macula-centered and optic nerve head-centric) of retinal fundus images captured from both the left and right eyes of the patients. The proposed multi-view stroke network, MVS-Net consists of multiple input layers, where high-level features are extracted from retinal fundus images using different state-of-the-art backbone models and then merged for the prediction of stroke. Our findings indicate that MVS-Net can detect stroke with $0.84$ AUC score contributing to the standardization of stroke assessment, where the network can be embedded into the software of a fundus camera and can be used to detect any stroke-related abnormalities in the patient's eyes. Hence, stroke assessment during emergency situations can be performed in the absence of neurologists.

\section*{Acknowledgements}
This study was supported by the Stroke-Data project under Business Finland Grant $3617/31/2019$. Authors would like to thank the research study nurses Riitta Laitala, Saara Haatanen, Matti Pasanen, and Tanja Kumpulainen, and research assistant Jari Paunonen for their contributions to data collection.

\section*{Author contributions}
A.D. contributed to conceptualization, methodology, software, formal analysis, data curation, and writing – original draft, M.H. contributed to supervising data analyses and editing the manuscript, J.P. contributed to conceptualization of the work, used methodology, outcomes curation, and editing the manuscript, P.H. contributed to conceptualization and methodology related to the fundus imaging, P.J. contributed to study planning, patient recruitment, patient data collection, funding, and conceptualization. All authors reviewed the manuscript.

\section*{Data availability statement}
The authors do not have permission to share data.

\bibliographystyle{IEEEtran}
\bibliography{IEEEtran}

\begin{thebibliography}{10}
\providecommand{\url}[1]{#1}
\csname url@samestyle\endcsname
\providecommand{\newblock}{\relax}
\providecommand{\bibinfo}[2]{#2}
\providecommand{\BIBentrySTDinterwordspacing}{\spaceskip=0pt\relax}
\providecommand{\BIBentryALTinterwordstretchfactor}{4}
\providecommand{\BIBentryALTinterwordspacing}{\spaceskip=\fontdimen2\font plus
\BIBentryALTinterwordstretchfactor\fontdimen3\font minus
  \fontdimen4\font\relax}
\providecommand{\BIBforeignlanguage}[2]{{%
\expandafter\ifx\csname l@#1\endcsname\relax
\typeout{** WARNING: IEEEtran.bst: No hyphenation pattern has been}%
\typeout{** loaded for the language `#1'. Using the pattern for}%
\typeout{** the default language instead.}%
\else
\language=\csname l@#1\endcsname
\fi
#2}}
\providecommand{\BIBdecl}{\relax}
\BIBdecl

\bibitem{PORTEGIES2016239}
M.~Portegies, P.~Koudstaal, and M.~Ikram, ``Chapter 14 - cerebrovascular
  disease,'' in \emph{Neuroepidemiology}, ser. Handbook of Clinical Neurology,
  M.~J. Aminoff, F.~Boller, and D.~F. Swaab, Eds.\hskip 1em plus 0.5em minus
  0.4em\relax Elsevier, 2016, vol. 138, pp. 239--261.

\bibitem{campbell2019ischaemic}
B.~C. Campbell, D.~A. De~Silva, M.~R. Macleod, S.~B. Coutts, L.~H. Schwamm,
  S.~M. Davis, and G.~A. Donnan, ``Ischaemic stroke,'' \emph{Nat. Rev. Dis.
  Primers}, vol.~5, no.~1, p.~70, 2019.

\bibitem{feigin2022world}
V.~L. Feigin, M.~Brainin, B.~Norrving, S.~Martins, R.~L. Sacco, W.~Hacke,
  M.~Fisher, J.~Pandian, and P.~Lindsay, ``World stroke organization (wso):
  global stroke fact sheet 2022,'' \emph{Int. J. Stroke.}, vol.~17, no.~1, pp.
  18--29, 2022.

\bibitem{doi:10.1161/01.STR.28.7.1507}
R.~L. Sacco, E.~J. Benjamin, J.~P. Broderick, M.~Dyken, J.~D. Easton, W.~M.
  Feinberg, L.~B. Goldstein, P.~B. Gorelick, G.~Howard, S.~J. Kittner, T.~A.
  Manolio, J.~P. Whisnant, and P.~A. Wolf, ``Risk factors,'' \emph{Stroke},
  vol.~28, no.~7, pp. 1507--1517, 1997.

\bibitem{ABSHER2002733}
J.~R. Absher, ``Cerebrovascular disease,'' in \emph{Encyclopedia of the Human
  Brain}, 2002, pp. 733--757.

\bibitem{andrade2012systematic}
S.~E. Andrade, L.~R. Harrold, J.~Tjia, S.~L. Cutrona, J.~S. Saczynski, K.~S.
  Dodd, R.~J. Goldberg, and J.~H. Gurwitz, ``A systematic review of validated
  methods for identifying cerebrovascular accident or transient ischemic attack
  using administrative data,'' \emph{Pharmacoepidemiol. Drug Saf.}, vol.~21,
  pp. 100--128, 2012.

\bibitem{SHARMA2017455}
V.~K. Sharma, ``Cerebrovascular disease,'' in \emph{International Encyclopedia
  of Public Health (Second Edition)}.\hskip 1em plus 0.5em minus 0.4em\relax
  Academic Press, 2017, pp. 455--470.

\bibitem{unnithan2022hemorrhagic}
A.~Unnithan and P.~Mehta, ``Hemorrhagic stroke.[updated 2022 feb 5],''
  \emph{StatPearls [Internet]. Treasure Island (FL): StatPearls Publishing},
  2022.

\bibitem{hatano1976experience}
S.~Hatano, ``Experience from a multicentre stroke register: a preliminary
  report,'' \emph{Bull World Health Organ.}, vol.~54, no.~5, p. 541, 1976.

\bibitem{amarenco2020transient}
P.~Amarenco, ``Transient ischemic attack,'' \emph{N Engl J Med.}, vol. 382,
  no.~20, pp. 1933--1941, 2020.

\bibitem{johnston2002transient}
S.~C. Johnston, ``Transient ischemic attack,'' \emph{N Engl J Med.}, vol. 347,
  no.~21, pp. 1687--1692, 2002.

\bibitem{vuong2015ocular}
L.~N. Vuong, P.~Thulasi, V.~Biousse, P.~Garza, D.~W. Wright, N.~J. Newman, and
  B.~B. Bruce, ``Ocular fundus photography of patients with focal neurologic
  deficits in an emergency department,'' \emph{Neurology}, vol.~85, no.~3, pp.
  256--262, 2015.

\bibitem{alijanpour2021different}
S.~Alijanpour, M.~Mostafazdeh-Bora, and A.~A. Ahangar, ``Different stroke
  scales; which scale or scales should be used?'' \emph{Caspian J. Intern.
  Med.}, vol.~12, no.~1, p.~1, 2021.

\bibitem{doi:10.1161/01.STR.30.8.1534}
S.~E. Kasner, J.~A. Chalela, J.~M. Luciano, B.~L. Cucchiara, E.~C. Raps, M.~L.
  McGarvey, M.~B. Conroy, and A.~R. Localio, ``Reliability and validity of
  estimating the nih stroke scale score from medical records,'' \emph{Stroke},
  vol.~30, no.~8, pp. 1534--1537, 1999.

\bibitem{doi:10.1111/j.1747-4949.2009.00294.x}
B.~C. Meyer and P.~D. Lyden, ``The modified national institutes of health
  stroke scale: its time has come,'' \emph{Int. J. Stroke}, vol.~4, no.~4, pp.
  267--273, 2009.

\bibitem{chaddad2023electroencephalography}
A.~Chaddad, Y.~Wu, R.~Kateb, and A.~Bouridane, ``Electroencephalography signal
  processing: A comprehensive review and analysis of methods and techniques,''
  \emph{Sensors}, vol.~23, no.~14, p. 6434, 2023.

\bibitem{birenbaum2011imaging}
D.~Birenbaum, L.~W. Bancroft, and G.~J. Felsberg, ``Imaging in acute stroke,''
  \emph{West. J. Emerg. Med.}, vol.~12, no.~1, p.~67, 2011.

\bibitem{10.1159/000362719}
M.~El-Koussy, G.~Schroth, C.~Brekenfeld, and M.~Arnold, ``{Imaging of Acute
  Ischemic Stroke},'' \emph{Eur. Neurol.}, vol.~72, no. 5-6, pp. 309--316,
  2014.

\bibitem{muir2006imaging}
K.~W. Muir, A.~Buchan, R.~von Kummer, J.~Rother, and J.-C. Baron, ``Imaging of
  acute stroke,'' \emph{Lancet Neurol.}, vol.~5, no.~9, pp. 755--768, 2006.

\bibitem{CHEUNG201789}
C.~Y. lui Cheung, M.~K. Ikram, C.~Chen, and T.~Y. Wong, ``Imaging retina to
  study dementia and stroke,'' \emph{Prog. Retin. Eye Res.}, vol.~57, pp.
  89--107, 2017.

\bibitem{mishra2022fundus}
C.~Mishra and K.~Tripathy, ``Fundus camera,'' in \emph{StatPearls [Internet]},
  2022.

\bibitem{jeena2019retina}
R.~Jeena, A.~Sukeshkumar, and K.~Mahadevan, ``Retina as a biomarker of
  stroke,'' in \emph{Comp. Aid. Interv. Diag. Clin. Med. Img.}, 2019, pp.
  219--226.

\bibitem{henderson2011hypertension}
A.~D. Henderson, B.~B. Bruce, N.~J. Newman, and V.~Biousse,
  ``Hypertension-related eye abnormalities and the risk of stroke,'' \emph{Rev.
  Neurol. Dis.}, vol.~8, no. 1-2, p.~1, 2011.

\bibitem{pula2017eyes}
J.~H. Pula and C.~A. Yuen, ``Eyes and stroke: the visual aspects of
  cerebrovascular disease,'' \emph{Stroke Vasc. Neurol.}, vol.~2, no.~4, 2017.

\bibitem{ZHU2022147}
G.~Zhu, H.~Chen, B.~Jiang, F.~Chen, Y.~Xie, and M.~Wintermark, ``Application of
  deep learning to ischemic and hemorrhagic stroke computed tomography and
  magnetic resonance imaging,'' \emph{Semin. Ultrasound CT MR}, vol.~43, no.~2,
  pp. 147--152, 2022, advances in Neuroradiology II : Artificial Intelligence.

\bibitem{pachade2022detection}
S.~Pachade, I.~Coronado, R.~Abdelkhaleq, J.~Yan, S.~Salazar-Marioni,
  A.~Jagolino, C.~Green, M.~Bahrainian, R.~Channa, S.~A. Sheth \emph{et~al.},
  ``Detection of stroke with retinal microvascular density and self-supervised
  learning using oct-a and fundus imaging,'' \emph{J. Clin. Med.}, vol.~11,
  no.~24, p. 7408, 2022.

\bibitem{jeena2021comparative}
R.~Jeena, G.~Shiny, A.~Sukesh~Kumar, and K.~Mahadevan, ``A comparative analysis
  of stroke diagnosis from retinal images using hand-crafted features and
  cnn,'' \emph{J. Intell. Fuzzy Syst.}, vol.~41, no.~5, pp. 5327--5335, 2021.

\bibitem{jeena2019stroke}
R.~Jeena, A.~Sukesh~Kumar, and K.~Mahadevan, ``Stroke diagnosis from retinal
  fundus images using multi texture analysis,'' \emph{J. Intell. Fuzzy Syst.},
  vol.~36, no.~3, pp. 2025--2032, 2019.

\bibitem{Lim_Lim_Xu_Ting_Wong_Lee_Hsu_2019}
G.~Lim, Z.~W. Lim, D.~Xu, D.~S. Ting, T.~Y. Wong, M.~L. Lee, and W.~Hsu,
  ``Feature isolation for hypothesis testing in retinal imaging: An ischemic
  stroke prediction case study,'' \emph{Proc. AAAI Conf. Artif. Intell.},
  vol.~33, no.~01, pp. 9510--9515, 2019.

\bibitem{coronado2021towards}
I.~Coronado, R.~Abdelkhaleq, J.~Yan, S.~S. Marioni, A.~Jagolino-Cole,
  R.~Channa, S.~Pachade, S.~A. Sheth, and L.~Giancardo, ``Towards stroke
  biomarkers on fundus retinal imaging: a comparison between vasculature
  embeddings and general purpose convolutional neural networks,'' in \emph{2021
  43rd Annu. Int. Conf. IEEE Eng. Med. Biol. Soc. (EMBC)}, 2021, pp.
  3873--3876.

\bibitem{diagnostics12071714}
N.~C. Khan, C.~Perera, E.~R. Dow, K.~M. Chen, V.~B. Mahajan, P.~Mruthyunjaya,
  D.~V. Do, T.~Leng, and D.~Myung, ``Predicting systemic health features from
  retinal fundus images using transfer-learning-based artificial intelligence
  models,'' \emph{Diagnostics}, vol.~12, no.~7, 2022.

\bibitem{YAN2021106}
X.~Yan, S.~Hu, Y.~Mao, Y.~Ye, and H.~Yu, ``Deep multi-view learning methods: A
  review,'' \emph{Neurocomputing}, vol. 448, pp. 106--129, 2021.

\bibitem{huang2017densely}
G.~{Huang}, Z.~{Liu}, L.~{Van Der Maaten}, and K.~Q. {Weinberger}, ``Densely
  connected convolutional networks,'' in \emph{IEEE Conf. Comput. Vision
  Pattern Recognit. (CVPR)}, 2017, pp. 2261--2269.

\bibitem{szegedy2016rethinking}
C.~{Szegedy}, V.~{Vanhoucke}, S.~{Ioffe}, J.~{Shlens}, and Z.~{Wojna},
  ``Rethinking the inception architecture for computer vision,'' in \emph{IEEE
  Conf. Comput. Vision Pattern Recognit. (CVPR)}, 2016, pp. 2818--2826.

\bibitem{he2016deep}
K.~{He}, X.~{Zhang}, S.~{Ren}, and J.~{Sun}, ``Deep residual learning for image
  recognition,'' in \emph{IEEE Conf. Comput. Vision Pattern Recognit. (CVPR)},
  2016, pp. 770--778.

\bibitem{simonyan2015a}
K.~Simonyan and A.~Zisserman, ``Very deep convolutional networks for
  large-scale image recognition,'' in \emph{3rd Int. Conf. Learn. Represent.
  (ICLR 2015)}, 2015, pp. 1--14.

\bibitem{8099678}
F.~Chollet, ``Xception: Deep learning with depthwise separable convolutions,''
  in \emph{2017 IEEE Conf. Comput. Vision Pattern Recognit. (CVPR)}, 2017, pp.
  1800--1807.

\bibitem{10.1007/978-3-030-58558-7_29}
A.~Kolesnikov, L.~Beyer, X.~Zhai, J.~Puigcerver, J.~Yung, S.~Gelly, and
  N.~Houlsby, ``Big transfer (bit): General visual representation learning,''
  in \emph{16th Eur. Conf. Comput. Vision (ECCV))}, 2020, p. 491–507.

\bibitem{5206848}
J.~Deng, W.~Dong, R.~Socher, L.-J. Li, K.~Li, and L.~Fei-Fei, ``Imagenet: A
  large-scale hierarchical image database,'' in \emph{2009 IEEE Conference on
  Computer Vision and Pattern Recognition}, 2009, pp. 248--255.

\bibitem{sacco2013updated}
R.~L. Sacco, S.~E. Kasner, J.~P. Broderick, L.~R. Caplan, J.~Connors,
  A.~Culebras, M.~S. Elkind, M.~G. George, A.~D. Hamdan, R.~T. Higashida
  \emph{et~al.}, ``An updated definition of stroke for the 21st century: a
  statement for healthcare professionals from the american heart
  association/american stroke association,'' \emph{Stroke}, vol.~44, no.~7, pp.
  2064--2089, 2013.

\end{thebibliography}

\end{document}